\documentclass[pra,showkeys,showpacs,superscriptaddress,twocolumn,nofootinbib,10pt]{revtex4-1}
\usepackage{graphicx}  
\usepackage{amsthm}
\usepackage{amsmath,amssymb,amsfonts,bbm,MnSymbol,mathrsfs,xfrac}   
\usepackage[breaklinks=true,colorlinks=true,linkcolor=blue,urlcolor=blue,citecolor=blue]{hyperref}
\usepackage{comment}
\usepackage{color}
\usepackage[makeroom]{cancel}

\renewcommand{\[}{\begin{equation}}
\renewcommand{\]}{\end{equation}}
\newcommand{\ket}[1]{|#1\rangle}
\newcommand{\bra}[1]{\langle#1|}

\newcommand{\ov}[1]{\overline{#1}}
\newcommand{\tr}{\mathrm{tr}}

\newcommand{\HS}{\mathcal{H}}

\newcommand{\R}{\hat{\rho}}
\newcommand{\de}{d\epsilon}

\newcommand{\be}{{\boldsymbol{\epsilon}}}

\newcommand{\vectorization}[1]{\mathrm{vec}{[#1]}}

\newtheorem{definition}{Theorem}

\newtheorem{theorem}[definition]{Theorem}

\begin{document}

\title{Simple expression for the quantum Fisher information matrix}

\author{Dominik \v{S}afr\'{a}nek}
\email{dsafrane@ucsc.edu}
\affiliation{SCIPP and Department of Physics, University of California, Santa Cruz, CA 95064, USA}

\date{\today}

\begin{abstract}
Quantum Fisher information matrix (QFIM) is a cornerstone of modern quantum metrology and quantum information geometry. Apart from optimal estimation, it finds applications in description of quantum speed limits, quantum criticality, quantum phase transitions, coherence, entanglement, and irreversibility. We derive a surprisingly simple formula for this quantity, which, unlike previously known general expression, does not require diagonalization of the density matrix, and is provably at least as efficient. With a minor modification, this formula can be used to compute QFIM for any finite-dimensional density matrix. Because of its simplicity, it could also shed more light on the quantum information geometry in general.
\end{abstract}

\maketitle

Quantum Fisher information matrix (QFIM) gives the ultimate precision bound on the estimation of parameters encoded in a quantum state.  This bound, called the quantum Cram\'er-Rao bound~\cite{helstrom1967minimum,Bures1969a,helstrom1976quantum,BraunsteinCaves1994a,Paris2009a,szczykulska2016multi}, gives the theoretical framework for maximizing sensitivity of new-era quantum detectors~\cite{Giovannetti2004a,Giovannetti2006a,zwierz2010general,Giovannetti2011a,Demkowicz2012a} such as recently improved~\cite{aasi2013enhanced,demkowicz2013fundamental} gravitational wave detector LIGO that confirmed the last missing piece in the Einstein's theory of relativity~\cite{abbott2016observation}. It is has been also used to find bounds on the optimal estimation of phases~\cite{Ballester2004a,Monras2006a,Aspachs2008a,demkowicz2009quantum,chin2012quantum,humphreys2013quantum,Nusran2014a,Sparaciari2015a,pezze2017optimal}, temperature~\cite{Monras2010a,correa2015individual,spedalieri2016thermal,hofer2017quantum}, space-time parameters~\cite{nation2009analogue,weinfurtner2011measurement,aspachs2010optimal,kish2017quantum,fink2017experimental}, magnetic fields~\cite{Wasilewski2010a,cai2013chemical,zhang2014fitting,nair2016far}, squeezing parameters~\cite{Milburn1994a,chiribella2006optimal,Gaiba2009a,benatti2011entanglement,safranek2016optimal}, time~\cite{zhang2013criterion,komar2014quantum}, and frequency~\cite{frowis2014optimal,boss2017quantum}. Apart from applications in quantum metrology, QFIM also gives speed limits on evolution of quantum states and quantum computation~\cite{margolus1998maximum,lloyd2000ultimate,taddei2013quantum,del2013quantum,binder2015quantacell,pires2016generalized,deffner2017quantum}, it quantifies coherence and entanglement~\cite{hauke2016measuring,girolami2017information,liu2017quantum}, and it provides bounds on irreversibility in open quantum systems~\cite{mancino2018geometrical}.

The Bures metric, which measures statistical distance between two infinitesimally close density matrices, is a different name for practically the same quantity. In fact, it has been shown~\cite{vsafranek2017discontinuities} that QFIM and the Bures metric are the same apart from single points, where QFIM suffers of removable discontinuities. QFIM and the Bures metric have been also used in the description of criticality and quantum phase transitions under the name of `fidelity susceptibility'. There they help to describe a sudden change of a quantum state when an external parameter such as temperature is varied~\cite{paraoanu1998bures,zanardi2007bures,venuti2007quantum,gu2010fidelity,banchi2014quantum,wu2016geometric,marzolino2017fisher}.

Considering the wide range of applicability of the QFIM, it is not a surprise that there has been a lot of effort in finding effective formulas for calculating it. We mention
Refs.~\cite{wootters1981statistical,hubner1992explicit,BraunsteinCaves1994a,slater1996quantum,dittmann1999explicit,sommers2003bures,Paris2009a,vsafranek2017discontinuities} that apply to quantum states in the density matrix formalism, and Refs.~\cite{Pinel2012a,Pinel2013b,Monras2013a,Gao2014a,Friis2015a,Safranek2015b,banchi2015quantum,nichols2017multiparameter,vsafranek2017calculating} that apply to Gaussian quantum states in the phase-space formalism.

Still, known expressions for the QFIM appear quite complicated, and the most general analytical expression requires diagonalizing the density matrix.

In this paper, we present a simple formula for the QFIM, which does not require any diagonalization, and applies to any finite-dimensional density matrix.

\section{Notation}

Lower indices will denote different matrices, while upper indices will denote elements of a matrix. \emph{Bar} as in $\ov{\R}$ will denote the complex conjugate, upper index $T$ as in $\R^T$ will denote transpose, and $\dag$ as in $\R^\dag$ will denote conjugate transpose. $\partial_i\equiv\partial_{\epsilon_i}$ denotes partial derivative with respect to $i$'th element of the vector of estimated parameters $\be=(\epsilon_1,\epsilon_2,\dots)$, $\hat{I}$ denotes the identity matrix, $\dim \HS$ denotes the dimension of the Hilbert space, $[\cdot,\cdot]$ denotes commutator, $\otimes$ denotes the Kronecker product, and $\vectorization{\cdot}$ denotes vectorization of a matrix, which is defined as a column vector constructed from columns of a matrix as
\[
A=\begin{pmatrix}
    a & b \\
    c & d \\
  \end{pmatrix}, \quad
  \vectorization{A}=\begin{pmatrix}
            a \\
            c \\
            b \\
            d \\
          \end{pmatrix}.
\]
We also drop index $\be$ showing the dependence of the density matrix on the vector of parameters, and write simply $\R\equiv\R_\be$.

The QFIM is defined~\cite{Paris2009a} as
\[\label{eq:qfi}
H^{ij}\equiv\tfrac{1}{2}\tr[(\hat{L}_i\hat{L}_j+\hat{L}_j\hat{L}_i)\R],
\]
where the symmetric logarithmic derivatives $\hat{L}_i$ are defined as operator solutions to equations
\[\label{eq:sld}
\tfrac{1}{2}\big(\hat{L}_i\R+\R\hat{L}_i\big)=\partial_i\R.
\]

\section{Results}
\begin{theorem}\label{thm1}
Let $\R$ be an invertible density matrix. The quantum Fisher information matrix can be computed as
\[\label{eq:QFIM}
H^{ij}=2\vectorization{\partial_i\R}^\dag\big(\ov{\R}\otimes \hat{I}+\hat{I}\otimes\R\big)^{-1}\vectorization{\partial_j\R},
\]
and the symmetric logarithmic derivatives as
\[\label{eq:SLDs}
\vectorization{\hat{L}_i}=2\big(\ov{\R}\otimes \hat{I}+\hat{I}\otimes\R\big)^{-1}\vectorization{\partial_i\R}.
\]
\end{theorem}

Using $\R^\dag=\R$, we can also rewrite the above equation as ${H^{ij}=2\vectorization{\partial_i\R^T}^T\big(\R^T\otimes \hat{I}+\hat{I}\otimes\R\big)^{-1}\vectorization{\partial_j\R}}$.

Since QFIM and the Bures metric are identical for invertible matrices~\cite{vsafranek2017discontinuities} (up to a multiplicative factor of $4$), and ${d\!\!\;\R}=\sum_i\partial_i\R{\!\ \de_i}$, the above result gives an expression for the infinitesimal Bures distance, $d_B^2(\R,\R+d\!\!\;\R)=\frac{1}{2}\vectorization{d\!\!\;\R}^\dag\big(\ov{\R}\otimes \hat{I}+\hat{I}\otimes\R\big)^{-1}\vectorization{d\!\!\;\R}$.

If $\R$ is invertible, then also $\mathfrak{M}\equiv\ov{\R}\otimes \hat{I}+\hat{I}\otimes\R$ is invertible. We can easily generalize the above result so that it also holds for non-invertible (singular) matrices $\R$, by using the result of Ref.~\cite{vsafranek2017discontinuities}. According to this paper, QFIM of any finite-dimensional density matrix can be computed as a limiting case of the QFIM of invertible density matrix $\R_\nu$:

\begin{theorem}\label{thm2}
Let $\R$ be any finite-dimensional density matrix, and $0<\nu<1$ a real parameter. We define invertible matrix $\R_\nu\equiv(1-\nu)\R+\frac{\nu}{\dim \HS}\hat{I}$. The quantum Fisher information matrix can be computed as a limit
\[\label{eq:not_invertible}
H^{ij}=2\lim_{\nu\rightarrow 0}\vectorization{\partial_i\R_\nu}^\dag\big(\ov{\R}_\nu\otimes \hat{I}+\hat{I}\otimes\R_\nu\big)^{-1}\vectorization{\partial_j\R_\nu}.
\]
\end{theorem}

\section{Proofs}
\begin{proof}(Theorem~\ref{thm1})

We are going to use the following identities~\cite{schacke2004kronecker,gilchrist2009vectorization}:
\begin{align}
\vectorization{ABC}&=(C^T\otimes A)\vectorization{B},\\
\tr[A^\dag B]&=\vectorization{A}^\dag\vectorization{B}.
\end{align}

We start with Eq.~\eqref{eq:sld}. This is a continuous Lyapunov equation, which can be expressed using vectorization and $\R=\R^\dag$ as
\[\label{eq:sldvec}
\begin{split}
\vectorization{\partial_i\R}&=\tfrac{1}{2}\vectorization{\hat{L}_i\R+\R\hat{L}_i}\\
&=\tfrac{1}{2}\vectorization{\hat{I}\hat{L}_i\R+\R\hat{L}_i\hat{I}}\\
&=\tfrac{1}{2}\big(\vectorization{\hat{I}\hat{L}_i\R}+\vectorization{\R\hat{L}_i\hat{I}}\big)\\
&=\tfrac{1}{2}\Big(\big(\R^T\otimes\hat{I}\big)\vectorization{\hat{L}_i}+\big(\hat{I}\otimes\R\big)\vectorization{\hat{L}_i}\Big)\\
&=\tfrac{1}{2}\big(\R^T\otimes\hat{I}+\hat{I}\otimes\R\big)\vectorization{\hat{L}_i}\\
&=\tfrac{1}{2}\big(\ov{\R}\otimes\hat{I}+\hat{I}\otimes\R\big)\vectorization{\hat{L}_i}.
\end{split}
\]
Assuming that $\R$ is invertible, solution to this equation can be written as
\[
\vectorization{\hat{L}_i}=2\big(\ov{\R}\otimes \hat{I}+\hat{I}\otimes\R\big)^{-1}\vectorization{\partial_i\R}.
\]

Using the above solution, the series of equalities follows:
\[
\begin{split}
H^{ij}&\equiv\tfrac{1}{2}\tr[(\hat{L}_i\hat{L}_j+\hat{L}_j\hat{L}_i)\R]\\
&=\tr[\partial_i\R\hat{L}_j]\\
&=\vectorization{(\partial_i\R)^\dag}^\dag\vectorization{\hat{L}_j}\\
&=\vectorization{\partial_i\R}^\dag\vectorization{\hat{L}_j}\\
&=2\vectorization{\partial_i\R}^\dag\big(\ov{\R}\otimes \hat{I}+\hat{I}\otimes\R\big)^{-1}\vectorization{\partial_j\R},
\end{split}
\]
which proves the Theorem.

Note: if $\R$ is not invertible, a solution to Eq.~\eqref{eq:sldvec} can be written as~\cite{ben2003generalized}
\[
\vectorization{\hat{L}_i}=2\big(\ov{\R}\otimes \hat{I}+\hat{I}\otimes\R\big)^+\vectorization{\partial_i\R},
\]
where upper index $+$ denotes the Moore-Penrose pseudoinverse~\cite{penrose1955generalized}. The QFIM can be calculated as
\[
H^{ij}=2\vectorization{\partial_i\R}^\dag\big(\ov{\R}\otimes \hat{I}+\hat{I}\otimes\R\big)^+\vectorization{\partial_j\R}.
\]
In case of invertible matrix, the inverse and the Moore-Penrose pseudoinverse coincide. This formula represents an alternative to Theorem~\ref{thm2}, however, since the pseudoinverse may be difficult to compute (some methods of construction can be found in Refs.~\cite{barnett1990matrices,petersen2008matrix}; or using Tikhonov regularization~\cite{van1996matrix}, $A^+=\lim_{\delta \searrow 0} (A^\dag A + \delta I)^{-1} A^\dag$), we do not stress this as our main result.
\end{proof}

\begin{proof} (Theorem~\ref{thm2})

Matrix $\R_\nu\equiv(1-\nu)\R+\frac{\nu}{\dim \HS}\hat{I}$ is invertible: this is because for each eigenvalue $0\leq\lambda_k\leq 1$ of the density matrix $\R$, the density matrix $\R_\nu$ has eigenvalue $\lambda_k^{(\nu)}=(1-\nu)\lambda_k+\frac{\nu}{\dim \HS}$ for which $0<\lambda_k^{(\nu)}<1$.

According to Ref.~\cite{vsafranek2017discontinuities}, the following Theorem holds:
\begin{theorem}\label{thm:regularization_procedure}
We define density matrix $\R_{\be,\nu}:=(1-\nu)\R_{\be}+\nu\R_{0}$, where $0<\nu<1$ is a real parameter and $\R_{0}$ is any $\be$-independent full-rank density matrix that is diagonal in the eigenbasis of the density matrix $\R_{\be}$. Then the resulting matrix $\R_{\be,\nu}$ is a full-rank matrix and
\[
H=\lim_{\nu\rightarrow 0}H(\R_{\be,\nu}).
\]
In finite-dimensional Hilbert spaces $\R_{0}$ can be defined as a multiple of identity, $\R_0=\frac{1}{\mathrm{dim}{\HS}}\hat{I}$.
\end{theorem}

Theorem~\ref{thm2} is therefore an application of the above Theorem, on Eq.~\eqref{eq:QFIM} for the finite-dimensional Hilbert spaces.
\end{proof}

\section{Discussion}
Definition of the QFIM, Eq.~\eqref{eq:qfi}, cannot be used directly, because symmetric logarithmic derivatives $L_i$ have to be found by solving Eq.~\eqref{eq:sld}. To the best of our knowledge, there have been only two known explicit expressions for the QFIM that can be directly applied to density matrices of any dimension. The first expression writes QFIM in terms of eigenvectors and eigenvalues of the density matrix $\R=\sum_{k}p_k|k\rangle\langle k|$, Refs.~\cite{hubner1992explicit,sommers2003bures,vsafranek2017discontinuities}, as
\[\label{QFI_multi}
H^{ij}=2\!\!\!\!\sum_{p_k+p_l> 0}\!\!\!\!\frac{\langle k|\partial_i\R|l\rangle\langle l|\partial_j\R|k\rangle}{p_k+p_l}.
\]
The second expression writes QFIM as an integral~\cite{Paris2009a},
\[\label{eq:integral_form}
H^{ij}=2\int_0^\infty\!\!\! dt\ \ \tr[e^{-\R t}\partial_i\R e^{-\R t}\partial_j\R].
\]

The first expression requires diagonalizing the density matrix, and the second is basis-independent, but requires exponentiation of the density matrix and computing the integral.\footnote{In fact, as we show in the Appendix, Eq.~\eqref{eq:QFIM} can be obtained by directly evaluating the intergral in Eq.~\eqref{eq:integral_form}.} Our formula, Eq.~\eqref{eq:QFIM}, represents an elegant alternative to the above expressions. It does not require diagonalizing the density matrix, nor any exponentiation and integration, but at the expense of computing the inverse of a relatively large matrix $\mathfrak{M}\equiv\ov{\R}\otimes \hat{I}+\hat{I}\otimes\R$. Finding this inverse may not be a problem for systems consisting of a few qubits, however, for larger systems one might have to employ efficient methods such as Cholesky decomposition~\cite{krishnamoorthy2013matrix}.

Notice that Eq.~\eqref{eq:QFIM} is valid for the density matrix written in any basis, which is its main advantage. For example, one can choose to work directly in the computational basis, which independent of estimated parameters. Choosing the basis to be the eigenbasis of the density matrix (which is usually parameter-dependent), matrix $\mathfrak{M}$ is diagonal and trivially inverted, and Eq.~\eqref{eq:QFIM} reduces to Eq.~\eqref{QFI_multi}. We can therefore conclude that Eq.~\eqref{QFI_multi} is a special case of our general expression, thus our general expression is at least as efficient in calculating the QFIM as Eq.~\eqref{QFI_multi}. It is important to note, that in terms of computational complexity it is probably not more efficient. The main advantages of our new expression therefore remains its matrix form, which makes it easy to implement, and the freedom to perform our computations in any basis we like.

In cases when we find diagonalizing the density matrix more convenient than inverting matrix $\mathfrak{M}$, but we want to stay in computational basis, we can choose a combined approach. Diagonalizing the density matrix is equivalent to finding a unitary decomposition of form $\R=UDU^\dag$, where $U$ is a unitary matrix consisting of eigenvectors of the density matrix, and $D$ is a diagonal matrix consisting of eigenvalues. Using $(A_1A_2)\otimes(B_1B_2)=(A_1\otimes B_1)(A_1\otimes B_1)$, we derive
\[
H^{ij}=2\vectorization{\partial_i\R}^\dag (\ov{U}\otimes U)\big(D\otimes \hat{I}+\hat{I}\otimes D\big)^{-1}(\ov{U}\otimes U)^\dag\vectorization{\partial_j\R}.
\]
Matrix in the middle is diagonal and therefore trivially inverted. Of course, this formula is nothing else than a matrix form of Eq.~\eqref{QFI_multi}.

Finally, let us consider a situation where parameters are encoded via unitary evolution $U(\be)=\exp(-i\sum_j\hat{K}_j\epsilon_j)$, as $\R=U(\be)\R_0U(\be)^\dag$. If Hermitian operators $\hat{K}_j$ commute, QFIM is independent of estimated parameters, and
\[
H^{ij}=2\vectorization{[\hat{K}_i,\R_0]}^\dag \big(\ov{\R_0}\otimes \hat{I}+\hat{I}\otimes \R_0\big)^{-1}\vectorization{[\hat{K}_j,\R_0]}.
\]
If $\R_0$ is not invertible, we can define invertible matrix $\R_{0\nu}\equiv(1-\nu)\R_0+\frac{\nu}{\dim \HS}\hat{I}$, and derive a special form of Eq.~\eqref{eq:not_invertible},
\[\label{eq:not_invertible_special}
H^{ij}\!=\!2\lim_{\nu\rightarrow 0}\vectorization{[\hat{K}_i,\R_{0\nu}]}^\dag\big(\ov{\R_{0\nu}}\otimes \hat{I}+\hat{I}\otimes\R_{0\nu}\big)^{-1}\!\!\vectorization{[\hat{K}_j,\R_{0\nu}]}.
\]

\section{Examples}

Here we give two examples with well-known results, to illustrate how the derived expressions can be used.

As the first example, we consider a simultaneous estimation of phase and noise. We start with initial state $\ket{\psi_{\theta}}=\frac{1}{\sqrt{2}}(\ket{0}+e^{-i \theta}\ket{1})$ which encodes the phase, but later deteriorates with white noise measured by parameter $\nu$. The resulting density matrix is
\[
\R=(1-\nu)\ket{\psi_{\theta}}\bra{\psi_{\theta}}+\frac{\nu}{2} \hat{I}
=\frac{1}{2}\begin{pmatrix}
 1 & (1-\nu)e^{i\theta} \\
 (1-\nu)e^{-i\theta} & 1 \\
\end{pmatrix}.
\]
We are going to compute limits on simultaneous estimation of parameters $\theta$ and $\nu$. We derive
\begin{gather}
\mathfrak{M}=
\begin{pmatrix}
1 &   e^{i\theta}\tfrac{\nu-1}{2} & e^{-i\theta}\tfrac{\nu-1}{2} & 0 \\
  e^{-i\theta}\tfrac{\nu-1}{2} &  1 & 0 & e^{-i\theta}\tfrac{\nu-1}{2} \\
 e^{i\theta}\tfrac{\nu-1}{2} & 0 &   1& e^{i\theta}\tfrac{\nu-1}{2} \\
 0 & e^{i\theta}\tfrac{\nu-1}{2} & e^{-i\theta}\tfrac{\nu-1}{2} &   1 \\
\end{pmatrix},\nonumber\\
\mathfrak{M}^{-1}=
\begin{pmatrix}
  1+\tfrac{1}{\nu(2-\nu)} &   e^{i\theta}\tfrac{\nu-1}{\nu(2-\nu)} & e^{-i\theta}\tfrac{\nu-1}{\nu(2-\nu)} & \tfrac{(\nu-1)^2}{\nu(2-\nu)} \\
  e^{-i\theta}\tfrac{\nu-1}{\nu(2-\nu)} &  1+\tfrac{1}{\nu(2-\nu)} & e^{-2i\theta}\tfrac{(\nu-1)^2}{\nu(2-\nu)} & e^{-i\theta}\tfrac{\nu-1}{\nu(2-\nu)} \\
 e^{i\theta}\tfrac{\nu-1}{\nu(2-\nu)} & e^{2i\theta}\tfrac{(\nu-1)^2}{\nu(2-\nu)} &   1+\tfrac{1}{\nu(2-\nu)} & e^{i\theta}\tfrac{\nu-1}{\nu(2-\nu)} \\
 \tfrac{(\nu-1)^2}{\nu(2-\nu)} & e^{i\theta}\tfrac{\nu-1}{\nu(2-\nu)} & e^{-i\theta}\tfrac{\nu-1}{\nu(2-\nu)} &   1+\tfrac{1}{\nu(2-\nu)} \\
\end{pmatrix},\nonumber\\
\vectorization{\partial_\theta\R}=\begin{pmatrix}
            0 \\
            -ie^{-i\theta}\tfrac{1-\nu}{2} \\
            ie^{i\theta}\tfrac{1-\nu}{2} \\
            0 \\
          \end{pmatrix},\
\vectorization{\partial_\nu\R}=\begin{pmatrix}
            0 \\
            -\tfrac{1}{2} e^{-i\theta}\\
            -\tfrac{1}{2} e^{i\theta} \\
            0 \\
          \end{pmatrix}.
\end{gather}
QFIM can be determined from Eq.~\eqref{eq:QFIM} as
\[
\begin{split}
H&=\begin{pmatrix}
 2\vectorization{\partial_\theta\R}^\dag\mathfrak{M}^{-1}\vectorization{\partial_\theta\R} &2\vectorization{\partial_\theta\R}^\dag\mathfrak{M}^{-1}\vectorization{\partial_\nu\R} \\
 2\vectorization{\partial_\nu\R}^\dag\mathfrak{M}^{-1}\vectorization{\partial_\theta\R} & 2\vectorization{\partial_\nu\R}^\dag\mathfrak{M}^{-1}\vectorization{\partial_\nu\R} \\
\end{pmatrix}\\
&=
\begin{pmatrix}
 (1-\nu)^2 & 0 \\
 0 & \tfrac{1}{\nu(2-\nu)} \\
\end{pmatrix}.
\end{split}
\]
For a single-shot experiment, quantum Cram\'er-Rao bound~\cite{Paris2009a,szczykulska2016multi} is a lower bound on the covariance matrix of estimators, and reads
\[
\mathrm{Cov}(\hat{\be})\geq H^{-1}.
\]
In other words, it says that matrix
\[
\mathrm{Cov}(\hat{\be})- H^{-1}=\begin{pmatrix}
 \mathrm{Var}(\hat{\theta})-\tfrac{1}{(1-\nu)^2} & \mathrm{Cov}(\hat{\theta},\hat{\nu}) \\
 \mathrm{Cov}(\hat{\theta},\hat{\nu}) & \mathrm{Var}(\hat{\nu})-\nu(2-\nu)  \\
\end{pmatrix}
\]
is positive semi-definite, which, according to Sylvester's criterion~\cite{prussing1986principal}, is equivalent to
\[
\begin{split}
\mathrm{Var}(\hat{\theta})&\geq\tfrac{1}{(1-\nu)^2},\\
\mathrm{Var}(\hat{\nu})&\geq\nu(2-\nu),\\
\mathrm{Cov}(\hat{\theta},\hat{\nu})^2&\leq \Big(\mathrm{Var}(\hat{\theta})-\tfrac{1}{(1-\nu)^2}\Big)\Big(\mathrm{Var}(\hat{\nu})-\nu(2-\nu)\Big).
\end{split}
\]
These inequalities show that error in estimation of $\theta$ and $\nu$ cannot fall below a certain threshold given by parameter $\nu$, and that correlation between the two parameters can be rather small, or non-existent.

From Eq.~\eqref{eq:SLDs}, we also calculate the symmetric logarithmic derivatives,
\begin{align}
L_{\theta}&=\begin{pmatrix}
 0 & -i(1-\nu)e^{i\theta} \\
 i(1-\nu)e^{-i\theta} & 0 \\
\end{pmatrix},\\
L_{\nu}&=\begin{pmatrix}
\tfrac{1-\nu}{(2-\nu)\nu} & e^{i\theta}\tfrac{1}{(2-\nu)\nu} \\
e^{-i\theta}\tfrac{1}{(2-\nu)\nu} & \tfrac{1-\nu}{(2-\nu)\nu} \\
\end{pmatrix}.
\end{align}
Eigenvectors of these operators give the optimal measurement bases that will allow us to satisfy the bounds written above, in the limit of many repetitions of the protocol~\cite{Paris2009a,szczykulska2016multi}. We find that both of these bases depends on parameter $\theta$. To optimally estimate parameter $\theta$ we should measure in basis
\[
B_{\theta}=\Big\{\tfrac{1}{\sqrt{2}}(-i e^{i\theta},1),\tfrac{1}{\sqrt{2}}(i e^{i\theta},1)\Big\},
\]
while to optimally estimate parameter $\nu$ we should measure in
\[
B_{\nu}=\Big\{\tfrac{1}{\sqrt{2}}(-e^{i\theta},1),\tfrac{1}{\sqrt{2}}(e^{i\theta},1)\Big\}.
\]

As the second example, we consider phase estimation using a maximally entangled two-qubit state. We start with initial state $\ket{\psi_0}=\frac{1}{\sqrt{2}}(\ket{0}\ket{0}+\ket{1}\ket{1})$, and assume that the phase is encoded in each of the qubits separately as $\ket{\psi_\theta}=\exp(-i\hat{K}\theta)\ket{\psi_0}$, where $\hat{K}\equiv\hat{N}\otimes \hat{I}+\hat{I}\otimes\hat{N}$ is the total number operator. $\hat{N}\otimes \hat{I}$ is the number operator acting on the first qubit, and $\hat{I}\otimes\hat{N}$ is the number operator acting on the second qubit respectively. Because the initial state is pure, in order to use Eq.~\eqref{eq:not_invertible_special} to calculate the QFIM, we have to define auxiliary density matrix $\R_{0\nu}\equiv(1-\nu)\ket{\psi_0}\bra{\psi_0}+\frac{\nu}{\dim \HS}\hat{I}$. We have
\begin{gather}
\hat{K}
=\begin{pmatrix}
0 &   0 & 0 & 0 \\
0&  1 & 0 & 0 \\
0 & 0 & 1 & 0 \\
0 & 0 & 0 &   2 \\
\end{pmatrix},\ \
\R_{0\nu}=\frac{1}{2}\begin{pmatrix}
1-\tfrac{\nu}{2} &   0 & 0 & 1-\nu \\
0&  \tfrac{\nu}{2} & 0 & 0 \\
0 & 0 & \tfrac{\nu}{2} & 0 \\
1-\nu & 0 & 0 &   1-\tfrac{\nu}{2} \\
\end{pmatrix},\nonumber\\
[\hat{K},\R_{0\nu}]=\begin{pmatrix}
0 &   0 & 0 & \nu-1 \\
0&  0 & 0 & 0 \\
0 & 0 & 0 & 0 \\
1-\nu & 0 & 0 &   0 \\
\end{pmatrix}.
\end{gather}
Matrices $\mathfrak{M}_{0\nu}=\big(\ov{\R_{0\nu}}\otimes \hat{I}+\hat{I}\otimes\R_{0\nu}\big)$ and $\mathfrak{M}_{0\nu}^{-1}$ are simple but rather large, so we will omit writing them here. QFIM is calculated from Eq.~\eqref{eq:not_invertible_special} as
\[
H=2\lim_{\nu\rightarrow 0}\left(\frac{4}{2-\nu}-4\nu\right)=4,
\]
which gives a lower bound on the error in estimation of phase,
\[
\mathrm{Var}(\hat{\theta})\geq\tfrac{1}{4}.
\]

\appendix

\section{Connection with the integral formula}
Here we show that our expression, Eq.~\eqref{eq:QFIM}, is the result of integration in the integral formula, Eq.~\eqref{eq:integral_form}.

We are going to use the following identities:
\begin{align}
\frac{d}{dt}(A\otimes B)&=\frac{d}{dt}A \otimes B+A\otimes \frac{d}{dt}B,\\
\tr[A^\dag B]&=\vectorization{A}^\dag\vectorization{B},\\
  (AB)\otimes(A'B')&=(A\otimes A')(B\otimes B'),\\
\vectorization{ABC}&=(C^T\otimes A)\vectorization{B},\\
(A\otimes B)^\dag&=A^\dag\otimes B^\dag.
\end{align}

Assuming that $\R$ is invertible, from the first identity we have
\[\label{eq:integral}
\frac{d}{dt}\Big(-\big(\ov{\R}\otimes \hat{I}+\hat{I}\otimes\R\big)^{-1}\big(e^{-\ov{\R}t}\otimes e^{-\R t}\big)\Big)=e^{-\ov{\R}t}\otimes e^{-\R t}.
\]

We start from the integral form of the quantum Fisher information matrix~\cite{Paris2009a},
\[
H^{ij}(\be)=2\int_0^\infty\!\!\! dt\ \ \tr[e^{-\R t}\partial_i\R e^{-\R t}\partial_j\R].
\]

The following equalities hold:
\begin{widetext}
\[
\begin{split}
H^{ij}(\be)&=2\int_0^\infty\!\!\! dt\ \ \vectorization{\big(e^{-\R t}\partial_i\R e^{-\R t}\big)^\dag}^\dag\vectorization{\partial_j\R}\\
&=2\int_0^\infty\!\!\! dt\ \ \vectorization{e^{-\R t}\partial_i\R e^{-\R t}}^\dag\vectorization{\partial_j\R}\\
&=2\int_0^\infty\!\!\! dt\ \ \bigg(\Big(\big(e^{-\R t}\big)^T\otimes e^{-\R t}\Big)\vectorization{\partial_i\R }\bigg)^\dag\vectorization{\partial_j\R}\\
&=2\int_0^\infty\!\!\! dt\ \ \vectorization{\partial_i\R }^\dag\bigg(\big(e^{-\R t}\big)^T\otimes e^{-\R t}\bigg)^\dag\vectorization{\partial_j\R}\\
&=2\int_0^\infty\!\!\! dt\ \ \vectorization{\partial_i\R }^\dag\Big(e^{-\ov{\R} t}\otimes e^{-\R t}\Big)\vectorization{\partial_j\R}\\
&=2\vectorization{\partial_i\R }^\dag\bigg(\int_0^\infty\!\!\! dt\ \ e^{-\ov{\R} t}\otimes e^{-\R t}\bigg)\vectorization{\partial_j\R}\\
&=2\vectorization{\partial_i\R }^\dag\Big[-\big(\ov{\R}\otimes \hat{I}+\hat{I}\otimes\R\big)^{-1}\big(e^{-\ov{\R}t}\otimes e^{-\R t}\big)\Big]_0^\infty\vectorization{\partial_j\R}\\
&=2\vectorization{\partial_i\R }^\dag\Big(-\big(\ov{\R}\otimes \hat{I}+\hat{I}\otimes\R\big)^{-1}\big(0-\hat{I}\otimes\hat{I}\big)\Big)\vectorization{\partial_j\R}\\
&=2\vectorization{\partial_i\R }^\dag\big(\ov{\R}\otimes \hat{I}+\hat{I}\otimes\R\big)^{-1}\vectorization{\partial_j\R},
\end{split}
\]
\end{widetext}
where we have used Eq.~\eqref{eq:integral} to compute the integral.

Similarly, starting from integral form of the symmetric logarithmic derivative~\cite{Paris2009a},
\[
\hat{L}_i=2\int_0^\infty\!\!\! dt\ \ e^{-\R t}\partial_i\R e^{-\R t},
\]
we derive
\[
\vectorization{\hat{L}_i}=2\big(\ov{\R}\otimes \hat{I}+\hat{I}\otimes\R\big)^{-1}\vectorization{\partial_i\R}.
\]

\bibliographystyle{apsrev4-1}
\bibliography{simpleBib}

\end{document}